\documentclass[english,submission]{eptcs}
\usepackage[T1]{fontenc}
\usepackage[latin9]{inputenc}
\usepackage{listings}
\usepackage{verbatim}
\usepackage{wrapfig}
\usepackage{graphicx}

\makeatletter
\usepackage{tikz}
\providecommand{\event}{PDMC 2011}
\def\titlerunning{Distributed MAP in the SpinJa Model Checker}
\def\authorrunning{Stefan Vijzelaar, Kees Verstoep, Wan Fokkink, and Henri Bal}

\@ifundefined{showcaptionsetup}{}{%
 \PassOptionsToPackage{caption=false}{subfig}}
\usepackage{subfig}
\makeatother

\usepackage{babel}

\begin{document}

\title{Distributed MAP in the SpinJa Model Checker}

\author{Stefan Vijzelaar, Kees Verstoep, Wan Fokkink, and Henri Bal\institute{VU University Amsterdam\\The Netherlands}\email{s.j.j.vijzelaar@vu.nl, c.verstoep@vu.nl, w.j.fokkink@vu.nl, h.e.bal@vu.nl}}
\maketitle
\begin{abstract}
Spin in Java (SpinJa) is an explicit state model checker for the Promela
modelling language also used by the SPIN model checker. Designed to
be extensible and reusable, the implementation of SpinJa follows a
layered approach in which each new layer extends the functionality
of the previous one. While SpinJa has preliminary support for shared-memory
model checking, it did not yet support distributed-memory model checking.
This tool paper presents a distributed implementation of a maximal
accepting predecessors (MAP) search algorithm on top of SpinJa. %
{}
\end{abstract}

\section{Introduction}

The SPIN model checker \cite{Holzmann2004} is arguably one of the
most popular explicit state model checkers to date. It can verify
models defined in its Promela modelling language for absence of deadlocks,
assertion violations, non-progress cycles and accepting cycles. This
has proved useful in many real-world applications by tracking down
problems and increasing reliability through modelling.

%
{}

Spin in Java (SpinJa) \cite{Jonge2010} is a model checker written
using object-oriented programming techniques. Its purpose is to be
easily extensible and reusable, while staying close to SPIN in terms
of usability, semantics, and performance. Whereas SPIN generates a
stand-alone verifier, SpinJa reuses its code through a library. %
{} Support ranges from language constructs up to verification algorithms.
In addition, the library can be easily extended due to its layered
structure. %
{}

A problem inherent to all model checkers, including SpinJa, is state
explosion. %
{} One way to deal with larger models is to distribute verification
over multiple machines in a distributed-memory cluster. The model
checker DiVinE \cite{Barnat2005} for example uses asynchronous communication
to achieve nearly linear speedups in verification \cite{Verstoep2009}.
While SpinJa has initial support for shared-memory model checking,
it lacks support for distributed-memory model checking.

This tool paper investigates how to extend SpinJa with support for
distributed-memory model checking. More specifically, it describes
the implementation of a distributed maximal accepting predecessors
(MAP) algorithm \cite{Barnat2005} in the generic layer of SpinJa.
The inherent extensibility of SpinJa should make this relatively easy,
but it is nevertheless interesting to see how this works out in practice.
It turns out that, except for a few problems, it is indeed possible
to reuse much of the SpinJa code-base in a distributed model checker.%
{}

Contributions to SpinJa cover both new features and bug fixes. New
features include meta-data support for state storage and remote referencing
of user-defined labels. A notable bug fix is the correct encoding
and decoding of states, which is used heavily in distributed model
checking. This code was not exercised thoroughly enough by SpinJa's
depth-first search. In addition to the SpinJa library, the implementation
contains communication facilities and termination detection.

The remainder of this paper is structured as follows. Section 2 gives
an overview of SpinJa, while section 3 gives an outline of the MAP
algorithm. The actual implementation of the algorithm in SpinJa is
discussed in section 4 and results are presented in section 5. The
paper is concluded in section 6.

\section{The SpinJa model checker}

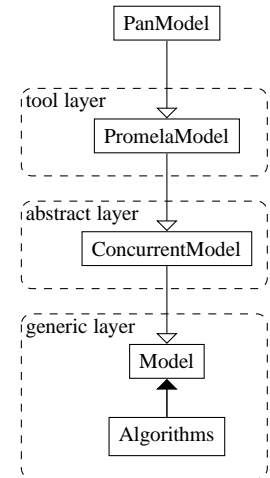
\begin{wrapfigure}{o}{0.3\columnwidth}%
\begin{centering}
\begin{tikzpicture}

\usetikzlibrary{arrows}
 
\draw (0,5.5) node (gm) [draw,rectangle,fill=white,font=\scriptsize] {PanModel};
\draw (0,4.0) node (pm) [draw,rectangle,fill=white,font=\scriptsize] {PromelaModel};
\draw (0,2.5) node (cm) [draw,rectangle,fill=white,font=\scriptsize] {ConcurrentModel};
\draw (0,1.0) node (md) [draw,rectangle,fill=white,font=\scriptsize] {Model};
\draw (0,0.0) node (va) [draw,rectangle,fill=white,font=\scriptsize] {Algorithms};

\draw [-triangle 90,fill=white] (gm) to (pm);
\draw [-triangle 90,fill=white] (pm) to (cm);
\draw [-triangle 90,fill=white] (cm) to (md);
\draw [-triangle 90,fill=black] (va) to (md);

\path (cm.west |- pm.north)+(-0.8,0.4) node (t1) {};
\path (pm.south -| cm.east)+(0.2,-0.3) node (t2) {};
\path[draw,rounded corners,dashed] (t1) rectangle (t2);
\draw (t1) node[anchor=north west,inner sep=1.5pt,font=\scriptsize] {tool layer};

\path (cm.west |- cm.north)+(-0.8,0.4) node (a1) {};
\path (cm.south -| cm.east)+(0.2,-0.3) node (a2) {};
\path[draw,rounded corners,dashed] (a1) rectangle (a2);
\draw (a1) node[anchor=north west,inner sep=1.5pt,font=\scriptsize] {abstract layer};

\path (cm.west |- md.north)+(-0.8,0.4) node (g1) {};
\path (va.south -| cm.east)+(0.2,-0.3) node (g2) {};
\path[draw,rounded corners,dashed] (g1) rectangle (g2);
\draw (g1) node[anchor=north west,inner sep=1.5pt,font=\scriptsize] {generic layer};

\end{tikzpicture}
\par\end{centering}

\caption{\label{fig:SpinJa-layers}SpinJa layers}
\end{wrapfigure}%
The Spin in Java (SpinJa) \cite{Jonge2010} model checker is designed
to be extensible and reusable. To achieve this goal, SpinJa uses a
layered design inspired by the framework of Mark Kattenbelt et al.\ \cite{Kattenbelt2007}.
SpinJa defines three layers: a generic, abstract and tool layer, see
Figure \ref{fig:SpinJa-layers}. Each layer extends the functionality
of the previous one. A parser is used to generate a compilable model,
extending the tool layer, from a Promela specification. Only the tool
layer is language specific. Algorithms, based on the generic layer,
are language agnostic.

The generic layer models the state space of the specification using
State and Transition objects. Simulation and verification algorithms
manipulate the state of the Model either directly by encoding and
decoding states, or indirectly by taking and undoing transitions.
Since algorithms are based on generic State and Transition objects,
they can easily share code, for example %
{}state hashing and storage methods. 

The abstract layer introduces concurrency between Models. The ConcurrentModel
defines one or more Processes which run concurrently%
{}; both ConcurrentModel and the Processes extend Model as defined by
the generic layer. Partial-order reduction is implemented at this
level and can limit the number of possible interleavings to %
{} improve performance.

The tool layer contains all language-specific facilities needed by
the model at run-time. The PromelaModel for example implements methods
to add process types and channels. A NeverClaimModel can be used to
verify the absence of unwanted behavior, and a RendezvousTransition
allows for rendezvous communication. This is the only language-specific
layer: the generic and abstract layer can be used by any tool layer
designed for a different language.

The parser uses JavaCC to translate a Promela specification to Java
source-code for a compilable model. The resulting PanModel will extend
the tool layer, using the run-time facilities it provides. The intermediate
compilation step mirrors the architecture of SPIN and helps to improve
performance by hard-coding information known at compile-time into
the model.

%
{}

\section{The MAP algorithm}

%
{}

\begin{wrapfigure}{r}{0.3\columnwidth}%
\begin{centering}
\begin{tikzpicture}[auto]

\draw (1.0,0.0) node (1) [draw,circle,fill=white] {1}
      (2.5,0.0) node (2) [draw,double,circle,fill=white] {2}
      (1.75,1.25) node (3) [draw,circle,fill=white] {3}
      (0.25,1.25) node (4) [draw,double,circle,fill=white] {4};

\draw [->] (-0.5,1.25) to (4);
\draw [->] (1) to (2);
\draw [->] (2) to node{} (3);
\draw [->] (3) to node{} (1);
\draw [->] (4) to node{} (3);

\end{tikzpicture}
\par\end{centering}

\caption{\label{fig:Accepting-cycle}Accepting cycle}
\end{wrapfigure}
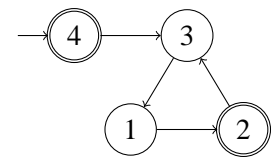%
Accepting cycles indicate the presence of an unwanted property. The
maximal accepting predecessors (MAP) algorithm \cite{Barnat2005}
is based on the knowledge that states in a cycle are their own predecessor.
Therefore, an accepting state is on a cycle if and only if it is its
own predecessor. Detecting this condition by keeping track of every
accepting predecessor for each visited state can significantly increase
memory requirements. Storage for a single state can increase linearly
with the number of accepting states in the model. The MAP algorithm
prevents this problem by storing only one accepting predecessor for
each state.

Storing a single accepting predecessor per state can however prevent
accepting cycles from being detected. The MAP algorithm assumes a
complete ordering of all accepting states in the state space and will
propagate to descendants only the accepting predecessor which is maximal
with respect to this ordering. However, when an accepting state on
an accepting cycle is being dominated by another accepting state outside
of the accepting cycle, it will not be detected as its own predecessor.
See Figure \ref{fig:Accepting-cycle}: the accepting state $2$ is
dominated by the accepting state $4$.

This problem is of no consequence if the dominating accepting state
is itself part of an accepting cycle. The goal of the algorithm is
to find only one accepting cycle, not all accepting cycles. One counter-example
is enough to disprove a property. However, if the dominating accepting
state is not part of an accepting cycle, then it needs to be removed
as an accepting state from the state space. This will allow the MAP
algorithm to run for a second iteration, without the state dominating
other accepting states. By iteratively changing accepting states to
non-accepting states, it becomes possible to detect any accepting
cycle in the state space. This process will continue until there are
no more accepting states in the state space, or until an accepting
cycle has been found.

In each iteration, information on accepting states is propagated to
all descendants. Nodes store information on the maximal accepting
predecessor encountered by states so far. The algorithm keeps track
of new accepting states it encounters, which can be removed from the
next iteration if they are not part of a cycle. Accepting states which
are dominated, and therefore won't be propagated, are not removed
at the end of an iteration, since it is possible for them to be part
of an undetected cycle.

\section{Implementing MAP in SpinJa}

In the implementation (available at \cite{SpinJa}), communication
for the MAP algorithm is handled by the Ibis Portability Layer (IPL)
\cite{Bal2010}. Ibis is a Java-based grid programming environment
allowing for highly efficient communication through its IPL communication
library. %
{} It also provides a convenient method for peer discovery: a registry
which can be used to coordinate the pool of workers for the MAP algorithm.
%
{}

The primary goal of the MAP implementation is to find cycles containing
accepting states. Accepting states are typically defined by Promela
never-claims, which in turn can be derived from linear-temporal-logic
(LTL) properties. Many of these properties, for example those from
the BEEM \cite{Pel'anek2007} database, reference specific labels
in the verification model. SpinJa supported accept, progress and end
labels in general, but only at the parser level. In particular, there
was no support for user-defined labels or for differentiating between
different instances of the already supported labels. (Promela defines
accept, progress and end labels by prefix.) Support for labels had
to be lifted to the run-time model of the verifier to allow for these
remote references in LTL properties.

Distributed MAP makes heavy use of SpinJa's encode and decode functions%
{}. Each worker is in a continuous loop of decoding a state, calculating
its successors and encoding these successors for shipment to other
workers. While SpinJa supports both breadth-first (BFS) and depth-first
(DFS) searches, mainly its DFS implementation had been thoroughly
tested. Instead of encoding and decoding model states, a sequential
DFS relies on the ability to take and undo specific transitions. It
turned out that the encode and decode functions worked incorrectly
when decoding a state sufficiently different from the current state.
(New processes when created during a decode would get an incorrect
process id.) A situation which does not occur in a DFS, but occurs
frequently in a BFS or distributed searches.

Transitions in SpinJa can store meta-data. This is for example useful
when performing a partial-order reduction to mark reduced transitions,
or when interleaving a never-claim to indicate which transitions cause
a context switch. States, however, do not support the storage of meta-data.
The DFS algorithm in SpinJa circumvents this problem by appending
meta-data to the state vector. However, this significantly increases
the stored state space, since the store does not differentiate between
the original state vector and the added meta-data.

In a MAP algorithm it is necessary to store the maximal accepting
predecessor propagated to states during execution: each state has
an associated MAP state. To support the storage of state meta-data,
SpinJa's ProbingHashTable had to be modified to allow explicit storage
of meta-data for the states it contains.

\begin{lstlisting}[caption={Distributed MAP iteration},label={lst:Distributed-MAP-iteration},basicstyle={\scriptsize},language=Java,numbers=left,numberstyle={\scriptsize},tabsize=4]
private boolean doIteration(byte iteration, SendPort[] senders) throws IOException {
	boolean flush = false;
	int shrink = 0;

	while (true) {
		while (!controlQueue.isEmpty()) {
			switch (processControl(controlQueue.remove(), senders, flush)) {
				case TERMINATE: return false;
				case ITERATE: return true;
				case FLUSH:
					workStack.clear();
					flush = true;
			}
		}

		while (!workStack.isEmpty() && controlQueue.isEmpty()) {
			shrink += processWork(workStack.pop(), iteration, senders);
			pollForMessages(flush);
		}

		while (!tokenQueue.isEmpty() && controlQueue.isEmpty() && workStack.isEmpty()) {
			processToken(tokenQueue.remove(), shrink == 0, senders);
		}

		if (controlQueue.isEmpty() && workStack.isEmpty() && tokenQueue.isEmpty()) {
			sendQueues(senders);
			blockForMessages(flush);
		} else {
			pollForMessages(flush);
		}
	}
}

\end{lstlisting}

At the heart of the distributed MAP algorithm lies the doIteration()
method shown in Listing \ref{lst:Distributed-MAP-iteration}. It represents
the main loop for each iteration of the MAP algorithm. Communication
is achieved using asynchronous message passing, similar to the DiVinE
\cite{Barnat2005} cluster model checker. The idea is to minimize
the influence of network latency and optimize parallelism by hashing
each generated state to a worker node and buffering it at the recipient.%
{}Termination is detected using a modified version of Safra's termination
detection algorithm \cite{Dijkstra1987}.

The controlQueue, stateQueue and tokenQueue store received messages
until they can be processed. These queues have an inherent priority
ordering: the controlQueue has the highest priority, followed by the
stateQueue, while the tokenQueue has the lowest priority. The controlQueue
stores messages which have an immediate influence on the control flow
of the algorithm. A FLUSH control message prevents any further states
from being processed, for example when a cycle has been found. The
ITERATE and TERMINATE control messages are used respectively to start
a new iteration or to finish the algorithm. The choice to ITERATE
or TERMINATE depends on whether any accepting states were dominated
in the last iteration. The stateQueue stores states which need to
have their successors calculated. Successor states are hashed to find
an owner node and are subsequently sent to that node. Finally the
tokenQueue stores tokens used for termination detection. These should
only be processed when the worker suspects the iteration to be finished:
in other words when the stateQueue is empty.

During the iterations the algorithm maintains two implicit sets: an
exclude-set and a shrink-set. The exclude-set contains accepting states
which should be ignored because they are not part of a cycle. This
set is necessary since SpinJa contains no facilities to remove a state
as accepting from the model. The shrink-set contains accepting states
which are candidates for removal in the next iteration. States are
both added and removed from this set during an iteration: newly encountered
accepting states are added to the set, while dominated accepting states
are removed. Should the iteration finish executing due to an ITERATE
control message, then the shrink-set is added to the exclude-set before
executing the next iteration.

%
{}

%
{}

\section{Results}

\begin{figure}[t]
\begin{centering}
\subfloat[reader\_writer (R=20, W=20, ERROR=1)]{\includegraphics[bb=0bp 0bp 504bp 340bp,clip,scale=0.45]{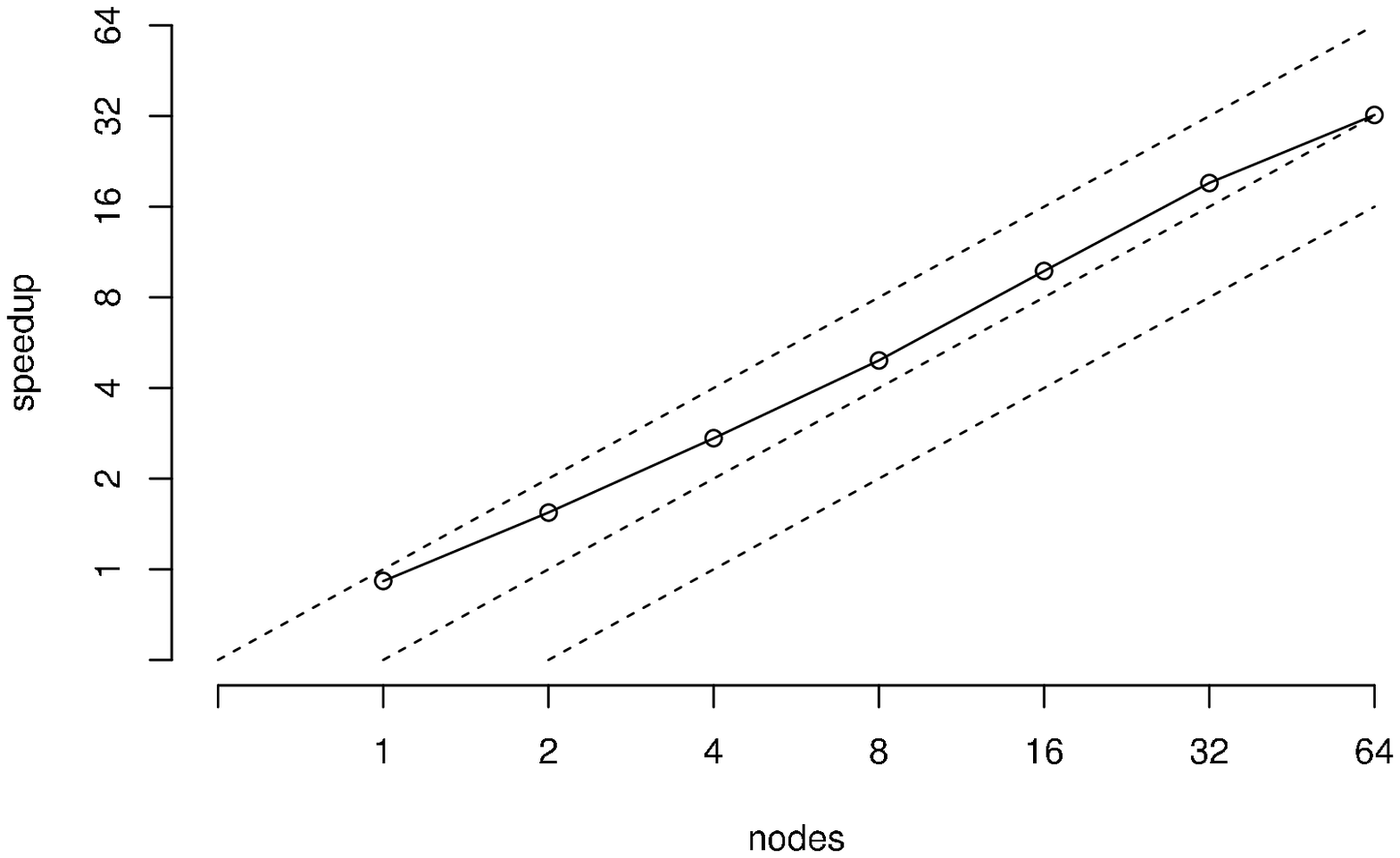}

}\subfloat[firewire\_link (N=5, APPLICATION=0)]{\includegraphics[bb=0bp 0bp 504bp 340bp,clip,scale=0.45]{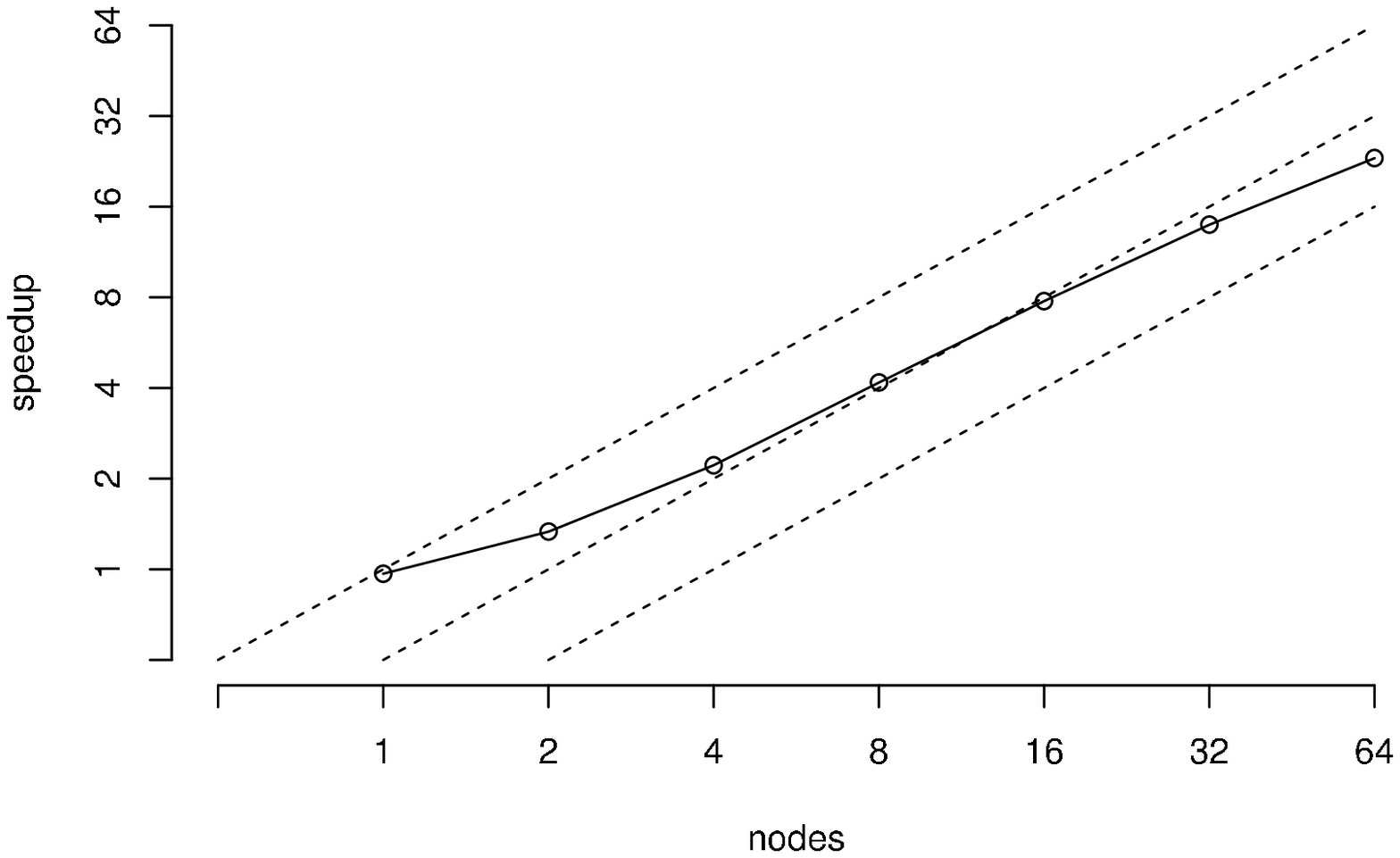}

}
\par\end{centering}

\caption{\label{fig:Scalability}Scalability}

\end{figure}
\begin{figure}[t]
\begin{centering}
\subfloat[lamport (N=5, ERROR=1)]{\includegraphics[bb=0bp 0bp 504bp 340bp,clip,scale=0.45]{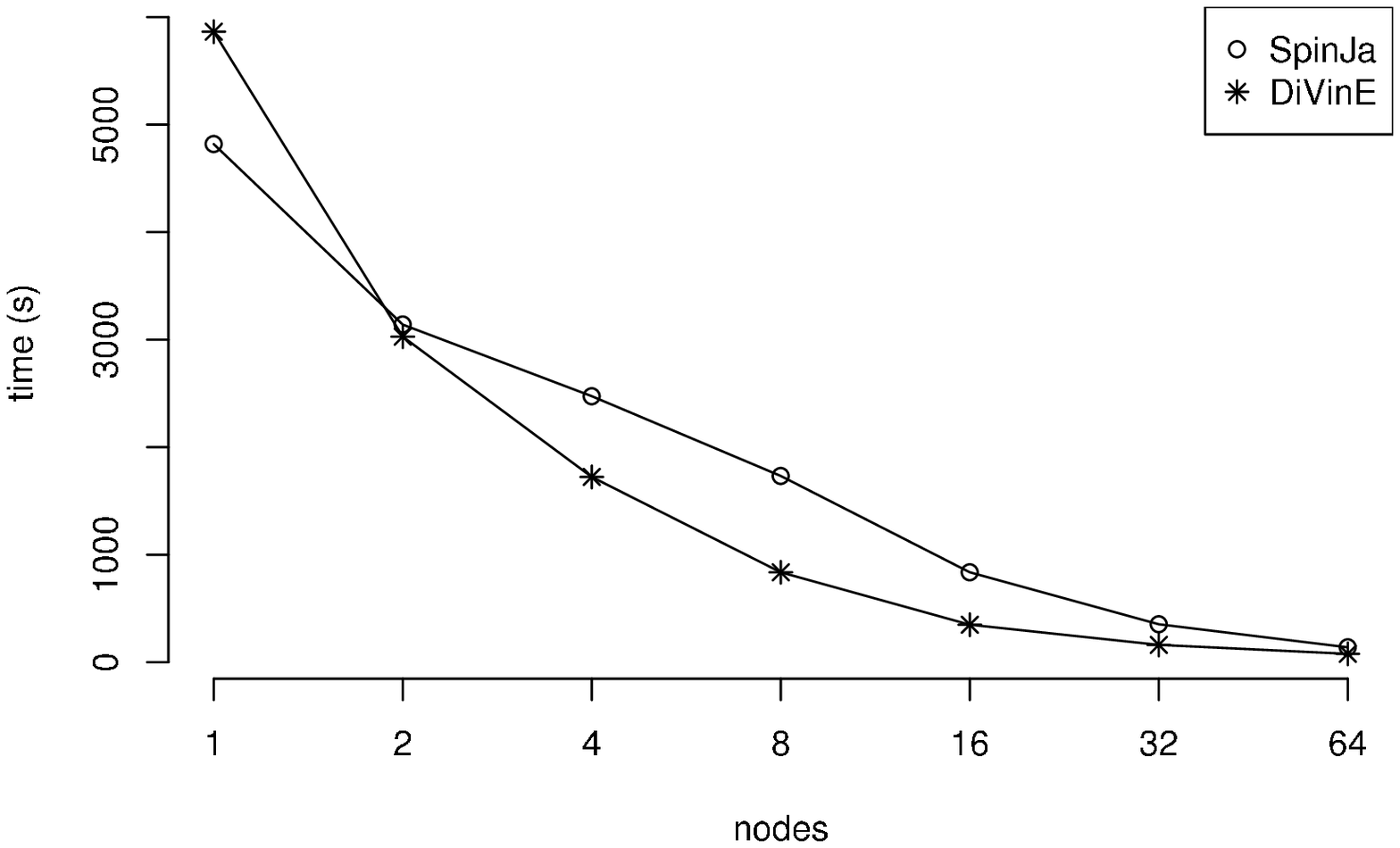}

}\subfloat[elevator (FLOORS=5, PERSONS=3)]{\includegraphics[bb=0bp 0bp 504bp 340bp,clip,scale=0.45]{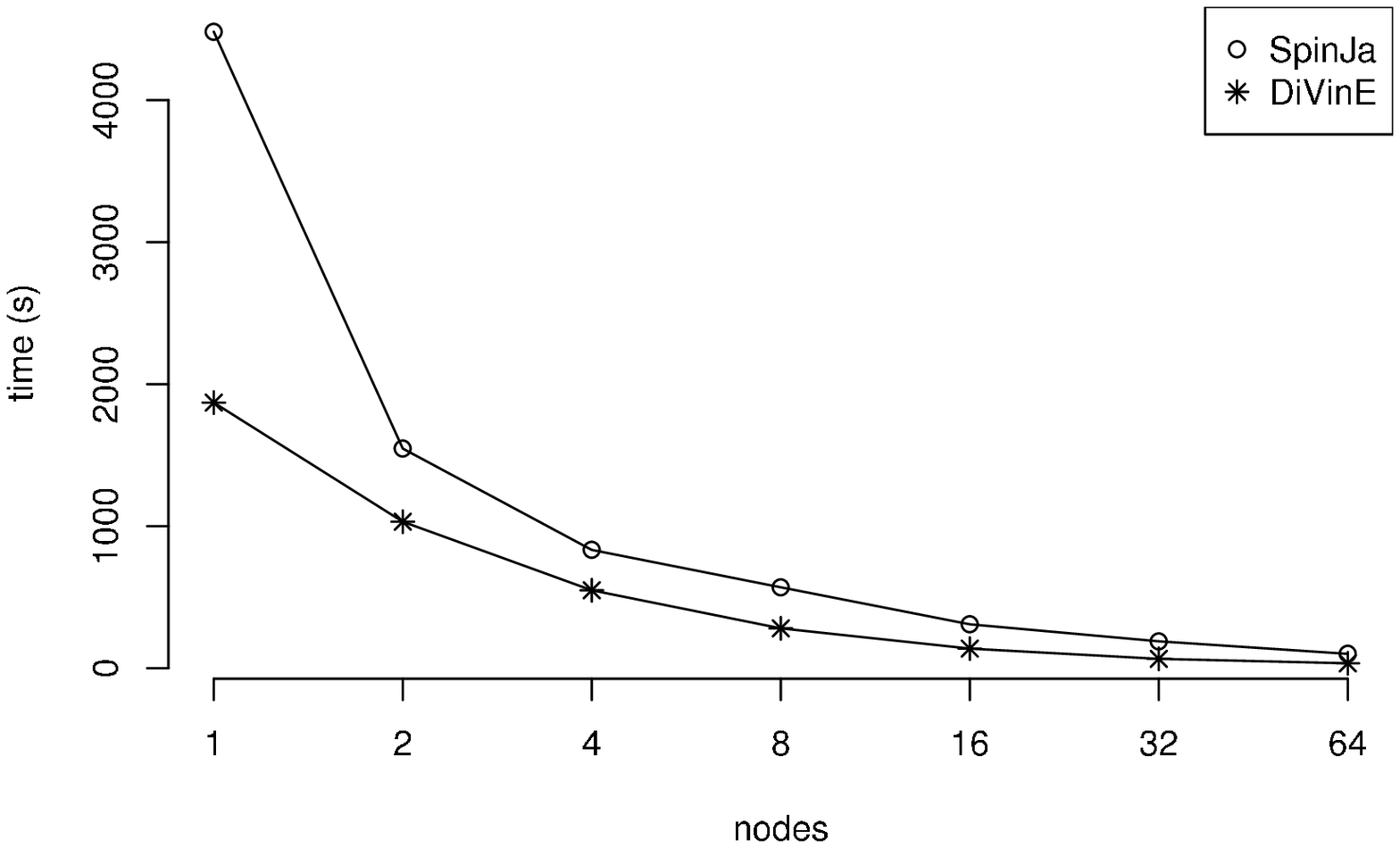}}
\par\end{centering}

\caption{\label{fig:Comparison}Comparison}

\end{figure}

In lieu of more thorough performance optimizations, the implementation
has been tested on the DAS-4 cluster \cite{DAS4} with four models
from the BEEM database \cite{Pel'anek2007}: reader\_writer, firewire\_link,
lamport and elevator. The sizes of these models range from approximately
600 million states (lamport), through 70 million (reader\_writer)
and 60 million states (firewire\_link), to 10 million states (elevator).
The graphs in Figure \ref{fig:Scalability} show the scalability of
reachability searches for two models. A comparison with the DiVinE
model checker is made in Figure \ref{fig:Comparison} using respectively
a reachability and a cycle search of the remaining two models.

Comparing the performance of the distributed MAP algorithm with sequential
DFS and BFS in SpinJa, shows the clear impact of communication overhead.
For example, a single node reachability search of the BEEM bakery.5
model, without network communication, takes DFS 29.1 seconds, BFS
34.3 seconds, and MAP 39.7 seconds. In comparison, enforcing network
communication (loop-back) for this single node in the MAP algorithm
increases its computation time to 50.1 seconds.

The scalability graphs in Figure \ref{fig:Scalability} show an average
efficiency above or around 50\%. The base used in these graphs is
the running time of a single-node DFS. The dotted lines respectively
indicate 100\%, 50\%, and 25\% efficiency. It can be seen, when scaling
up from a single node, that the algorithm looses efficiency because
of an increased amount of non-local network communication. At higher
node counts the initial setup of network connections becomes increasingly
significant as connection setup time increases, and the total running
time decreases.

Comparing the performance of the distributed MAP algorithm with DiVinE
in Figure \ref{fig:Comparison} shows that the results are of a similar
order. In this case the graph uses the total running time instead
of scalability, since the nested DFS algorithm used for cycle searches
in SpinJa is not optimal (due to the lack of generic meta-data storage).
Scalability of DiVinE is more predictable than that of the presented
MAP algorithm, even though results are generally close.

Clearly some performance optimizations are still required. The improvements
described by Verstoep et al. \cite{Verstoep2009} should prove valuable:
combining messages, automatic tuning of the polling rate, prioritisation
of timing critical communication and avoiding congestion while flushing
message queues.

\section{Conclusion}

The goal of this paper has been to explore the flexibility of the
SpinJa model checker by extending it to distributed-memory model checking.
Implementing a distributed MAP algorithm in SpinJa turns out to be
relatively easy: except for some missing features, most of the SpinJa
library can be reused as is. The exceptions are the framework for
sequential search algorithms in the generic layer, and the partial-order
reduction algorithm which is not equipped to handle a distributed
state space. The interface of the generic layer, based on states and
transitions, proves to be a good starting point for model checkers
wanting to support the Promela modelling language.

In addition to the facilities provided by the SpinJa library, a distributed
algorithm also requires communication facilities to distribute states.
Communication is explicitly programmed throughout the implementation
presented in this paper. An interesting alternative would be to extend
SpinJa's framework to abstract away from those communication primitives.
Aspects like termination detection could be reused between algorithms.
This avenue has been explored in parallel to the implementation presented
in this paper. SpinJa has been successfully employed to generate state
spaces for the distributed graph algorithm framework HipG\cite{Krepska2011}. 

It will be interesting to see how the SpinJa implementation compares
to the more generic HipG solution, when verifying large-scale models
on a large distributed-memory cluster like the DAS-4 \cite{DAS4}.
Especially after the performance improvements described in \cite{Verstoep2009}
have been applied to SpinJa. %
{}
\begin{description}
\item [{Acknowledgements}] Thanks go to Elzbieta Krepska for identifying
various bugs in SpinJa.
\end{description}
\bibliographystyle{eptcs}
\bibliography{pdmc11}

\end{document}